\documentclass[a4paper,10pt]{article}

\usepackage{geometry}
\makeatletter\@addtoreset{chapter}{part}\makeatother%
\usepackage[dvipdfmx]{graphicx}
\usepackage[normal,oneline]{caption2}
\usepackage{url}
\usepackage[affil-it]{authblk}
\usepackage{textgreek}
\usepackage{xcolor}
\usepackage{bm}
\usepackage{appendix}

\begin{document}

\providecommand{\keywords}[1]
{
  \small	
  \textbf{\textit{Keywords---}} #1
}

\title{\bf Calibration of X-Ray Source of a Powder Diffractometer\\
and Radiation Test of Silicon Microstrip Detectors}



\author[a]{E.~Fokitis}
\author[b]{T.~Geralis}
\author[a]{S.~Maltezos}
\author[c]{N.~Vodinas}

\affil[a]{National Technical University of Athens (NTUA), Physics Department}
\affil[b]{National Centre of Scientific Research ``Demokritos'' (NRCPS)}
\affil[c]{Ministry of Education and Religious Affairs, Greece}

\date{}
\maketitle

\pagenumbering{arabic}

\newcommand{\beq}{\begin{equation}}
\newcommand{\eeq}{\end{equation}}

\newcommand{\ben}{\begin{eqnarray}}
\newcommand{\een}{\end{eqnarray}}

\begin{abstract}
\emph{Abstract.}
A flexible apparatus for calibration of the absolute flux at the focal plane of the X-ray Source of a Powder Diffractometer, based on a fast scintillator counter, is presented. The measured fluxes, depending on the high voltage on the X-ray tube, were at the range 200 - 400 MHz, while an uncertainty in the flux of the order of 5$\%$ has been estimated. We also applied this calibration for radiation hardness study of a multichannel silicon microstrip X-Ray detector.  

\noindent
\end{abstract} \hspace{10pt}

{\keywords{Powder Diffractometer, X-Rays, Silicon microstrips}}

\section{Introduction}

This work deals with the assembly of a detector prototype for the calibration of the intensity of a linear focus X-ray source of 1.5~kW electrical power, and the operational test of silicon microstrip detectors; the latter are used in the goniometric circle of a development diffractometer of Debye-Scherrer type. The beam is produced in a monochromatic mode after dispersion with monochromator of bent crystal. The calibration of the X-ray source will be necessary for the radiation tests of silicon detectors since they are expected to indicate (possible) changes due to radiation (Radiation Damage) and will give an estimate of their useful time. The knowledge of the relative peaks in the Debye-Scherrer spectra allows us to determine the relative strength of various crystal phases in each sample, but for an absolute estimation of these, the absolute flux of the X-ray source is necessary and is usually required in a X-ray refractometer.
The intrinsic calibration difficulty for the X-ray beam relates to the counting rates which are of the order of 1~GHz, and pose well-known difficulties in coping with these. In the past, methods used for calibration employ absorbers with calibrated material named ``Lupolen'' (platelet of Polyethylene), and recently use of single or double slits.

The design presented in this work relates to a fast detector system with high efficiency in detecting X-rays in the energy range of 5-20~keV. We present here  the counting rates for the spectral emission Cu-K$_{\mathrm{a1}}$ as a function of the source current and high voltage. The appropriate correction coefficients for the dead time and geometrical parameters of the detection system are given in this work. 

The experimental design and configuration is presented in Section 2. In Section 3, the experimental measurements and the data analysis methodology are analyzed. In Section 4, an analytical comparison of our detector with a commercial detector is performed. Finally, in Section 5, we present a radiation hardness test of microstrip silicon detectors, while in Section 6 the conclusions and prospects are discussed.

\section{Design of the detection system}
\label{System}

For the choice of the appropriate detector we studied the characteristics of detectors with high efficiency in X-rays in the range of 8~keV, such as silicon detectors of PIN or Avalanche Photodiodes, and organic or inorganic scintillators. The first category needs cooling and continuous temperature control due to the strong dependence of the signal and background on the temperature, and because of problems arising from the exposure to large radiation fluxes. The scintillators (organic and inorganic) are in general easier to use, with exception of the hygroscopic ones, and especially these   having extremely small de-excitation time,τ (for example, the BaF$_2$ scintillator has $\tau\approx0.6-0.8$~ns, and the plastic scintillators have $\tau\approx 1-3$~ns). We discuss below the appropriate scintillator for the detection of 8 keV X-rays together with characteristics in response times, light yield and scintillation emission spectrum induced by the X-ray absorption. 

\begin{figure}[!ht]
\centering
\begin{minipage}[b]{0.4\linewidth}
\centering
\includegraphics[scale=0.3]{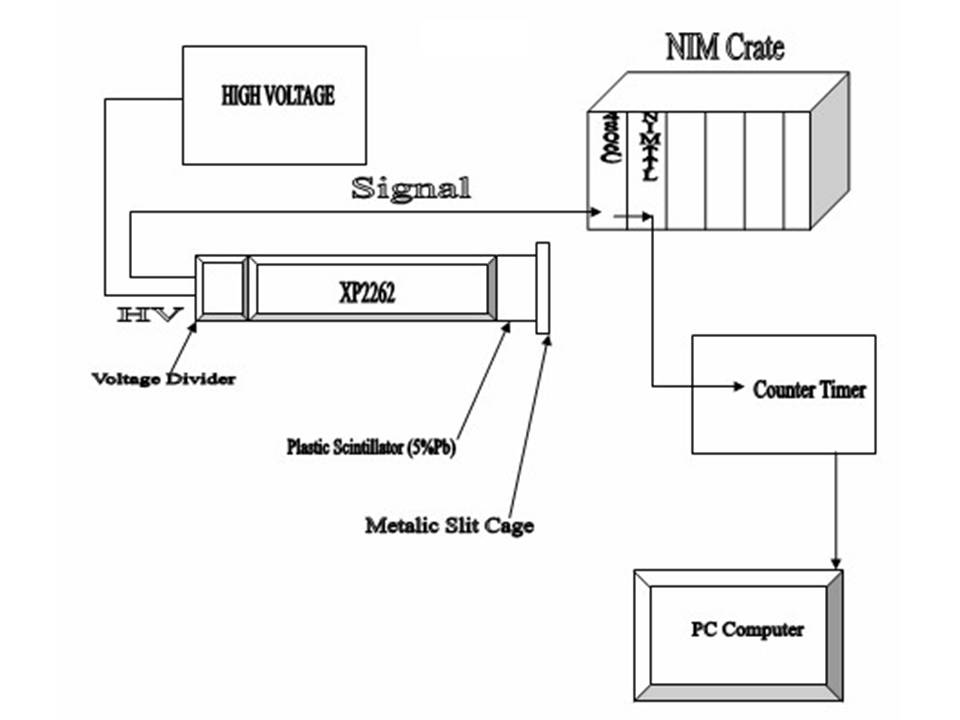}  
\caption[]{The experimental setup of the calibration apparatus.}
\label{setup}
\end{minipage}
\qquad
\begin{minipage}[b]{0.36\linewidth}
\centering
\includegraphics[scale=0.3]{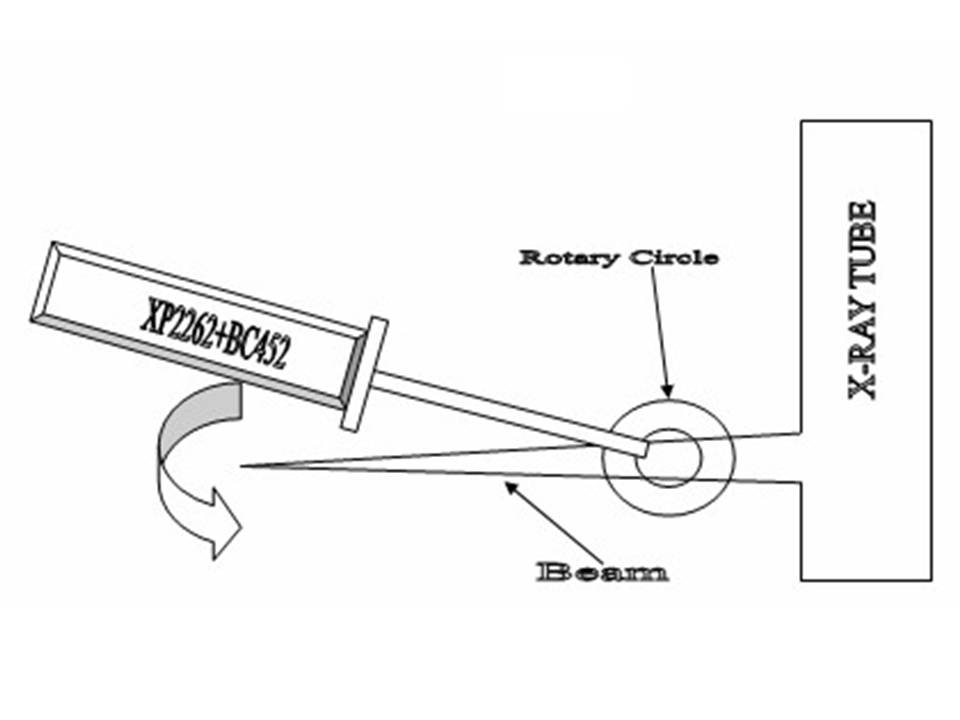}
\caption[]{Schematics indicating the use of the X-ray detector prototype.}
\label{detector}
\end{minipage}
\end{figure}

Scintillators, such as Nai(Tl), which is customarily used for the X-ray detection, due to its high efficiency and the large number of scintillation photons, is not recommended for large radiation fluxes because of its large de-excitation time. The BaF$_2$, among the fastest scintillators presents an additional requirement: the need of an optical filter for isolating the fast component $(\lambda<220)$~nm from the slow one $\lambda > 220$~ nm. One of its disadvantages which appear also in certain fast scintillators is the small number of produced scintillation photons in the visible range for each X-ray photon; this number is considerably smaller in comparison with the yield from a plastic scintillator.  Thus, the scintillator of model BC452 of Bicron which was selected for our application, is doped with Lead (5$\%$ Pb), has a small absorption length (large absorption coefficient: $\mu=4.91$~cm$^{-1}$ at 20~keV), and an extremely fast response (de-excitation time, $\tau \sim 2.2$τ~ns). The number of emitted photons corresponds to $15\%$ of the photon yield of NaI(Tl), and thus, the pulse produced needs no further amplification. The emission spectrum of BC452, as given by the manufacturer, extends in the range 400-480~nm, and peaks at around 424~nm. After an extensive search, the photomultiplier tube (PMT) of XP2262 of Photonis \cite{PMT} was selected to be combined with the selected scintillator BC452. The spectral response of the photocathode convoluted with the borosilicate window transmittance matches nicely the emission spectrum of the scintillator. We observe a peak combined yield at $\lambda=420$~nm, and a rather narrow FWHM $\approx 30$~nm. The narrow spectral bandwidth of our detection system indicates the small optical noise which would arise from small and ultimately un-avoidable light leaks.

\subsection{Experimental setup}

\begin{figure}[!ht]
\centering
\includegraphics[scale=0.38]{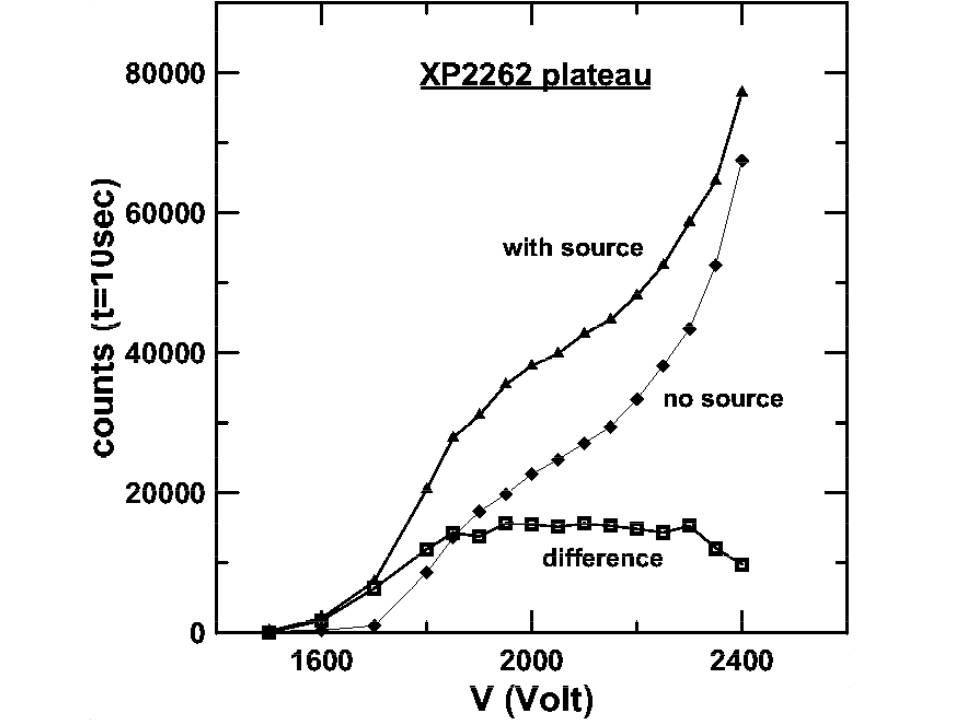}  
\caption{The PMT XP2262 plateau using the X ray tube. We see measurements with the source (triangular data points), without source , i.e. dark current (square black data points) and the difference of the two (square white points).} 
\label{plateau}
\end{figure}

The scintillator and the PMT were optically coupled and the system was made light-tight. Both were surrounded by a metallic shield. A housing-cover was attached to the scintillator side in such a way as to allow the insertion of a variable width slit (see Fig. \ref{setup} and Fig. \ref{detector}). The PMT was connected with an appropriate HV distribution circuit which allowed the possibility of optimum control of the voltage between the first two dynodes. For the X-ray flux measurement, we selected the method of operation in a ``photon counting mode''. Thus, the analogue pulse from the PMT is input to the discriminator, model 4806C of LeCroy, and this output is input to a NIM to TTL converter. Finally, the TTL pulse is recorded by a homemade fast ($100~MHz$) counter-timer \cite{Maltezos}. The latter had the capability to be controlled by a PC computer, which also records the number of counts in a predefined time gate (t=10~s). Before the flux measurements, we studied the PMT plateau curve versus High Voltage and the dark current behaviour seen in Fig. \ref{plateau}. 

The selected-optimum High Voltage was 2000~V and the threshold of the discriminator was set at -30~mV. Next, the detector was at a distance of 26.8~cm from the capillary tube and its axis was in the X-ray beam direction. To reduce the detected rate of X-ray photons, a metallic slit of width $d=40\pm2~\mathrm{\mu}$m was placed in front of the scintillator entrance. We next describe a set of measurements done to study, first the characteristics of the detector, namely its efficiency, dead time, and secondly, the calibration of the X-ray tube flux for various High Voltage values, and tube currents.

\section{Experimental measurements and data analysis}
\subsection{Detector dead time and calibration}

Setting the High Voltage in the X-ray tube at 16, 20, 25 and 30~kV, the X-ray fluxes were recorded for X-ray tube currents varying from 5~mA up to the maximum allowed value for each High Voltage selected, i.e. 16,20, 25 and 30~mA, respectively. The step in the tube current increment was 1~mA. For each of the above settings the dark counts of the detector were recorded before and after each measurement. The values of the dark current corresponding counting rate were of the order of 2-5~kHz. This value is consistent with the PMT manufacturer's specifications.
For large X-ray flux values, the well-known pile-up effect causes significant deviations of the measured rates from the true ones, and above certain limit, we have the phenomenon of paralysis \cite{Foglio}, \cite{Libert_1}, \cite{Libert_2}, \cite{Muller}, \cite{Usman}. The true beam intensity $R$ may be written as a linear function of the X-ray tube current $I$:

\beq
R(I)=aI+\beta
\label{rate}
\eeq

where, $R(I)$ is the counting rate of X-rays (counts/s) and $I$ is given in [mA].  The coefficients to be determined, $a$ and $\beta$, are called the calibration parameters. Due to the finite dead time, let $\tau$, the recorded rate of X-rays, $m$, is: 

\beq
m=R(I)e^{-R\tau}=(aI+\beta)e^{-(aI+\beta)\tau}
\label{rec_rate}
\eeq
 
\begin{figure}[!ht]
\centering
\includegraphics[scale=0.60]{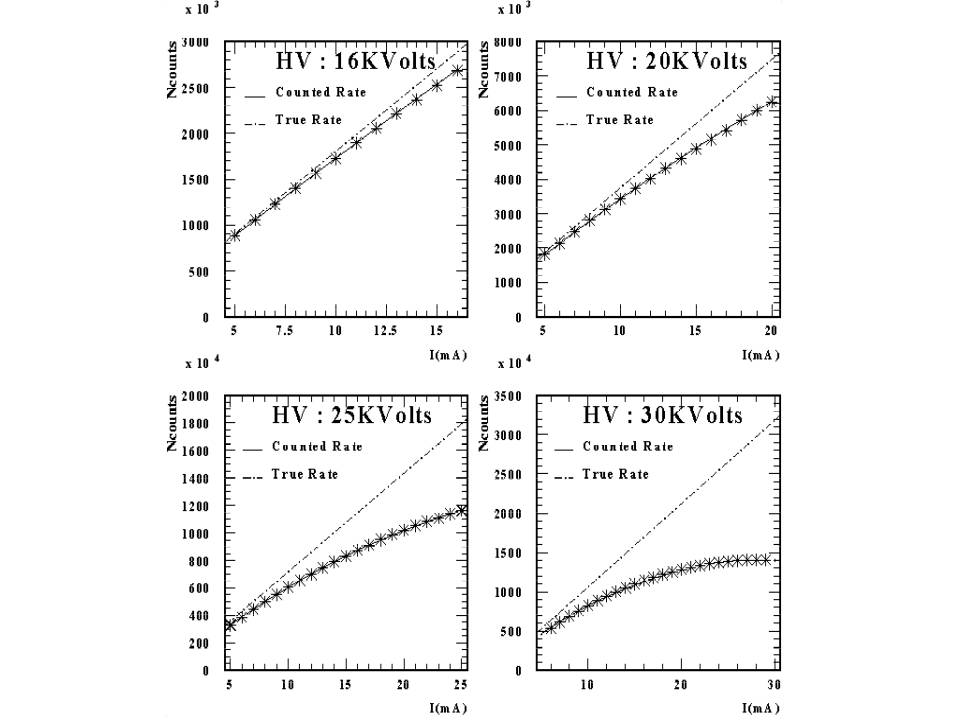}  
\caption{Measured counting rate as a function of X-ray tube current. The asterisks represent the data points of the measurements, the continuous line shows the fitted curve, and the dotted line indicate the true rate, as it is obtained by using the parameters $a$ and $\beta$.} 
\label{rates}
\end{figure}

The dead time of the detector is also considered as a parameter to be determined at the same time with the parameters $a$ and $\beta$ by non-linear $\chi^2$ fitting method. In Fig. \ref{rates}, we observe the four different sets of measurements for X-ray tube HV: 16, 20, 25 and 30~kV. The continuous line is the result of the fit and the dotted line represents the actual counting rate $R$.

We observe, as seen in this figure, the higher the counting rates the stronger is the pile-up effect. In addition, we should note that for the case HV=30~kV, the fit fails if we include more than the 15 first points on the curve. This effect is due to the loss of linearity in the PMT performance; this may be due to reasons such as large space charges  appearing between its dynodes and the induced deviation from linear change of the dynodes potentials \cite{Manfredi}. Despite these effects, and up to HV around 40~kV, a good number of data points, in small current values can be used in order to obtain calibration for all allowed HV values in the range of 10-40~kV. 
A criterion for the correctness of this calibration method can be the agreement of the dead time values of the detector, as they are extracted independently from our four data sets. 

\begin{figure}[!ht]
\centering
\includegraphics[scale=0.45]{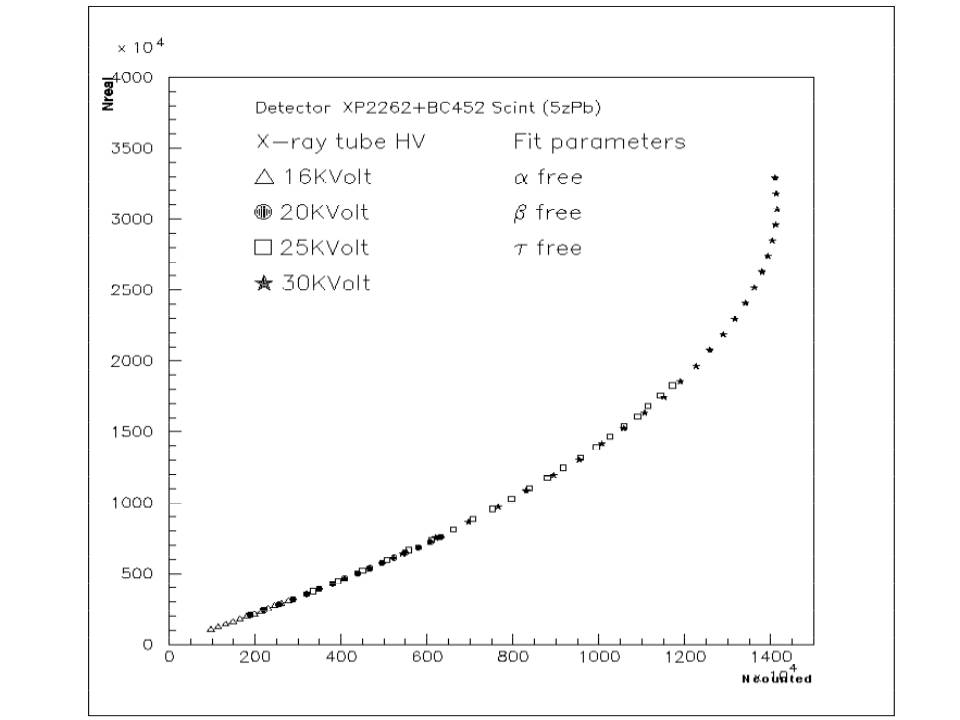}  
\caption{True rate of X-ray photons ($\mathrm{N_{real}}$) as a function of the recorded rate ($\mathrm{N_{counted}}$).} 
\label{true}
\end{figure}

We used four combinations of the fitting parameters, $a$, $\beta$ and $\tau$. In two cases, alternatively, we had set constant values, $\beta=0$ and $\tau=24.3$~ns observing small fluctuations in the determined free parameters. The parameter $\beta$ was statistically equal to zero in all the cases, which shows a ``healthy'' behaviour of the fitting procedure. The parameter $a$ differed, in the four cases studied, from the average value at most by $1.5\%$, $1.7\%$, $0.6\%$, and $2.5\%$, respectively. The largest among these fluctuations ($2.5\%$) is used, as the worst systematic error. It should be noted that the statistical sample in each measurement was quite large and, as a result, the above quoted systematic error may be considered very close to the total error in the determination of above parameters. In Table \ref{dead_time}, the dead time and the parameter $a_{\mathrm{HV}}$ for the different HV settings are given. The obtained results were statistically consistent between each other. 
The value of the dead time of the detector which we accept is the statistical weighted average, using as weight the inverse of the square of the error,that is, $\tau=24.3\pm 0.1$~ns.
In Fig. \ref{true}, the true rate as a function of the measured photon rate is shown for all the parameters assumed free, where an excellent degree of agreement, concerning the continuity of measurements for the various HV values of the X-ray tube, is seen.

\begin{table}[!h]
\centering
\begin{tabular}{|c|c|c|c|c|}
\hline 
High Voltage~[kV] &  Dead time~[ns] & $a_{\mathrm{HV}}[\mathrm{mA}^{-1}\mathrm{s}^{-1}]\times10^3$ & $\chi ^2/\mathrm{dof}$  \\ 
\hline 
16             & $29.6\pm3.1$    &   $181.5\pm5$  &  8/9\\ 
\hline  
20             & $22.6\pm1.1$    &   $372.2\pm9.3$  &  7/13\\ 
\hline 
25             & $24.4\pm0.1$    &   $719.1\pm18$  &  24/18\\ 
\hline 
30             & $24.3\pm0.1$    &   $1076\pm27$  &  17/12\\ 
\hline
\end{tabular} 
\caption{The dead time and the parameter $a_{\mathrm{HV}}$, obtained by the fitting procedure, for the four different values of X-ray tube High-Voltage.} 
\label{dead_time}
\end{table}

\subsection{Comparative tests with SZINTIX detector}
The above methodology was applied also for a similar commercial detector system, model SZINTIX \cite{SZINTIX}, which is based in a NaI(Tl) scintillator and has quite slower electronic units (amplifier, discriminator etc). The reasons which have led us to this additional test were, on the one hand, to confirm the validity of the method, and, on the other hand, to determine the efficiency of the detector ``XP2262/BC452'' in lower photon rates, as it is given that the SZINTIX detector has efficiency $100\%$ at 8~keV. The dead time of the detector SZINTIX, for the corresponding cases 16, 20, 25, and 30~kV, with use of Aluminum absorber of thicknesses $(2.8\pm2.6)$, $(7.6\pm2.1)$, $(5.9\pm0.8)$, and $(5.4\pm 0.2)$
$~\mathrm{\mu}$m, giving an average value $\tau=(5.42\pm0.19)~\mathrm{\mu}$s. The detection efficiency of XP2262/BC452 at 8~keV, using similar geometry for the two detectors, was found to be: $A_{\mathrm{det}}=1.00\pm0.02$. This means that the efficiency of our detector system is nearly $100\%$.

\subsection{Efficiency of the detector in $Cu-K_{a1}$ X-rays}
The efficiency of the detector is defined as the ratio of the experimentally recorded number of the X-rays to the actually incident ones. The inefficiency of the detector is related to the number of expected photons, within the spectral window of sensitivity of the PMT, produced by each X-ray. According to our estimations the number of such photons at 8 keV is around 46 for the scintillator BC452. The number of photons incident on the PMT depends on the angular distribution of the scintillation and the details of the geometry of the scintillator. Therefore, there should be an appropriate simulation in order to get an accurate estimation of the efficiency. However, assuming that around $15\%$ of the scintillation photons are incident on the PMT photocathode, we obtain that around 7 photons per incident X-ray are recorded. This should be multiplied by the average collection efficiency $(\sim0.95)$ and quantum efficiency of the PMT $(\sim 0.25)$. Thus, a guessed number for the detector efficiency, taking into account these considerations, might give a value of around 0.81. However, for higher energy X-ray photons the efficiency is expected to be significantly improved.

\subsection{Correction of counting rate}
Using the above data, and the corresponding analysis, it was possible to determine the dead time of our detector system, which was $\tau=(24.3\pm0.1)$~ns. Furthermore, by solving Eq. \ref{rec_rate} for $R$, we may determine the true X-rays rate. The solution of this transcendental equation is done numerically, by the Newton-Raphson method. The correction factor, $f(m)$, as a non linear function of the measured counting rate, $m$, and its parameterization is, $f(m)=4.08\times {{10}^{-29}}{{m}^{4}}-7.47\times {{10}^{-22}}{{m}^{3}}+6.17\times {{10}^{-15}}{{m}^{2}}+1.21\times {{10}^{-8}}m+1$. Its value ranges from 1.0 to 1.9. The real rate is then calculated by the formula, $R=f(m)m$.

\section{Determination of absolute intensity of the beam}
It must be noted that the parameter $a$ corresponds to X-rays rate for a slit of width $d=40$~$\mathrm{\mu}$m at a distance of $r=26.8$~cm from the capillary tube. The slit location was chosen so that it is near the peak of the normal distribution of the beam cross-section (more precisely, at $0.53\sigma$ from the distribution peak).
The parameter $a$ suffices for the determination of the real X-rays photon rate, when the HV and the current $I$ of the X-ray tube is known. At the present time, the value of $a$ was determined only in the four cases discussed. In the future, on systematic study is expected to give the value of $a$ for continuous variation of HV in the range of 10-40~kV. In such a case, an additional correction must be applied, regarding the detection efficiency by dividing $R$ with $A_{\mathrm{det}}$. Then the intensity of the beam, in [Hz], passing through the slit, with $I$ expressed in [mA], is:

\beq
I_{\mathrm{slit}}=\frac{aI}{A_{\mathrm{det}}}
\eeq

\begin{figure}[!ht]
\centering
\includegraphics[scale=0.45]{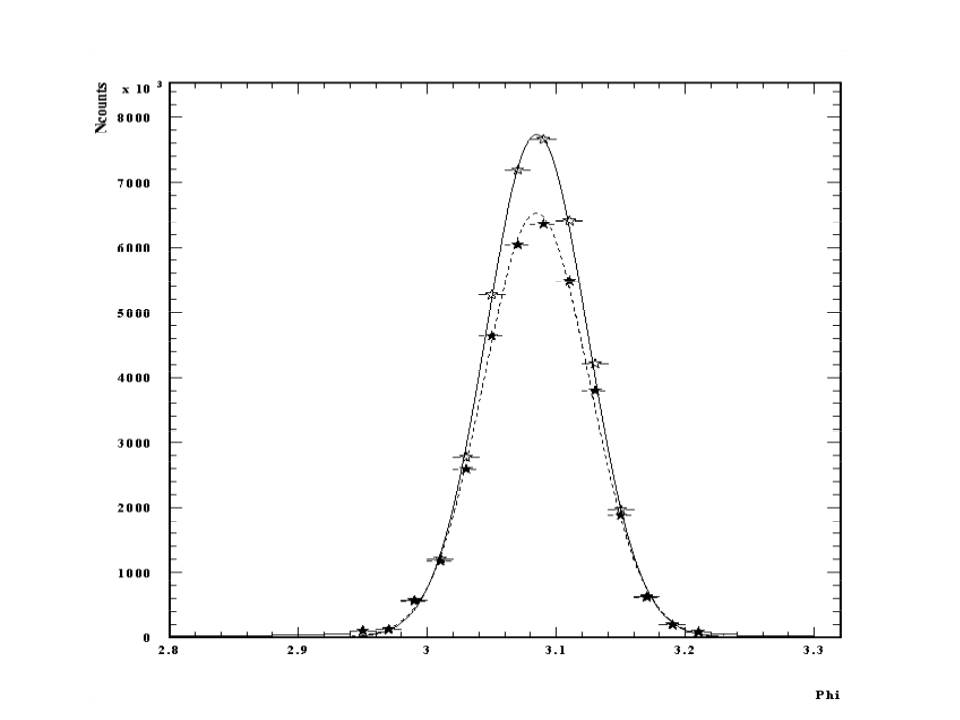}  
\caption{The recorded X-ray beam profile in [counts/s] as a function angle $\phi$ (dashed line obtained from the fitting of the data shown with dark circles), while the solid line represents the data after dead time corrections. } 
\label{beam_profile}
\end{figure}

For the determination of the absolute beam intensity we applied the following method: 
keeping the values of X-ray tube HV and current constant (25~kV and 10~mA, respectively) we measured the flux, at various positions, of the beam intensity profile with an angular step of $0.02^o$, which corresponds to a length variation of 93.5~$\mathrm{\mu}$m at distance 26.8~cm from the bent crystal to the detector slit of width 40~$\mathrm{\mu}$m. In Fig.\ref{beam_profile}, we observe the results of this scanning. By performing a $\chi^2$ fit with a Gaussian distribution, we find that the mean value which corresponds to angle $<\phi>=3.085^0$ and standard deviation, $\sigma_{\phi}=0.039^o$ or FWHM$=0.091^o$. As a result, the absolute beam intensity in [Hz] is given by:

\beq
{{I}_{\mathrm {tot}}}=\frac{\sum{{{N}_{i}}}}{{{N}_{{{i}_{\max }}}}}\frac{{{D}_{\mathrm{step}}}}{{{d}_{\mathrm {slit}}}}\frac{1}{{{A}_{\det }}}{{\varepsilon }_{\mathrm {corr}}}N_{{{i}_{\max }}}^{'}=\frac{\sum{{{N}_{i}}}}{{{N}_{{{i}_{\max }}}}}\frac{{{D}_{\mathrm {step}}}}{{{d}_{\mathrm {slit}}}}\frac{1}{{{A}_{\det }}}{{a}}I~(\mathrm {mA})
\eeq

obtaining,

\beq
{{I}_{\mathrm{tot}}}=12.47\frac{1}{{{A}_{\det }}}{{\varepsilon }_{\mathrm{corr}}}N_{{{i}_{\max }}}^{'}=12.47\frac{1}{{{A}_{\det }}}aI~(\mathrm{mA})
\eeq

where $N_i$ is the corrected recorded rate of the $i^{th}$ element, is the recorded  rate at the element of the maximum,  $D_{\mathrm{strip}}=$93.5~$\mathrm{\mu}$m is the spatial step;  $D_{\mathrm{strip}}=40~\mathrm{\mu}$m is the slit width, $\varepsilon_{\mathrm{corr}}$ is the correction factor and $N_{{{i}_{\max }}}^{'}$ is the recorded rate at the maximum of the combination of HV and current to the evaluated one (for example, 30~kV, 15~mA). The overall relative error of $I_{\mathrm{tot}}$ is:

\beq
\frac{\sigma_{I_{\mathrm{tot}}}}{I_{\mathrm{tot}}}=\frac{{{\sigma }_{{{d}_{\mathrm{slit}}}}}\oplus {{\sigma }_{{{A}_{\det }}}}\oplus {{\sigma }_{\mathrm{N}}}}{{{I}_{\mathrm{tot}}}}=0.06
\eeq

The main contribution in the overall error comes from the measurement of the slit width. More accurate measurement of this parameter with metrological method is expected to give significant improvement to the beam's absolute intensity accuracy.

Two examples of computation of the total beam intensity follow:\\

${{I}_{\mathrm{tot}}}(\mathrm{30~kV},\mathrm{15 mA})=5.33\times 2.34\times 1.00\times 1.489\times 10990000=\mathrm{(204 }\!\!\pm\!\!\mathrm{ 12) ~MHz}$

${{I}_{\mathrm {tot}}}(\mathrm{30~kV},\mathrm {30 mA})=5.33\times 2.34\times 1.00\times 1076325\times 30=\mathrm {(402 }\!\!\pm\!\!\mathrm { 24) ~MHz}$\\

For the radiation hardness test we selected the settings HV=25~kV and $I=10$~mA. The measured rate was 6.045~MHz while the corrected one was 7.250~MHz at $0.53\sigma$ and on the peak was 7.800~MHz. 

\section{Radiation hardness of microstrip silicon detector}
For the radiation hardness studies we used a microstrip detector, which comprises of 2000 microstrip elements of width $30~\mu$m and heigth 1~cm, while the distance of the centers of two successive strips is $50~\mu$m. We placed the center of the microstrip detector at the position corresponding to the peak of the intensity of the X-Ray beam as determined above. According to the results of the previous sections, based on the slit used, the rate is 7.8~MHz. Converting this rate, which corresponds to an irradiated area $40~\mu$m by 1.5~cm, to the microstrip detector pixel size of $30~\mu$m by 1~cm, we obtain an irradiation rate of 3.9~MHz. 

Additionally, based on the FWHM of the beam, we can find a corresponding spatial beam width equal to $420~\mu\mathrm{m}$ (for a distance of 269~mm from the capillary tube - center of circumference of sweeping of beam). Since the distribution of the beam intensity is nearly Gaussian, the width of the detector which accepts high X-ray flux is expected to extend up to $\pm 5\sigma$ or about 4~FWHM we get $1680~\mu\mathrm{m}$. In this range, about 34 silicon microstrips are exposed or 17 microstrips on each side of the central microstrip. In addition, the exposure time in this photoflux was 12 hours. After the irradiation of the microstrip detectors,  measurements of leakage currents were carried out (Institute of Microelectronics, Demokritos) in order to investigate if there are aging effects which could be due to creation of surface states in the interface Si-SiO$_2$ induced by the X-rays irradiation. In the center of the detector, the leakage current at potential -100~V (full depletion) was found to be on the  average around two or three times larger in comparison to the current before irradiation. Three microstrips on the left of the 100$^{th}$ (central) microstrip had large leakage current (larger than 20~nA). As we are departing from the center, the leakage current tends to the values recorded before irradiation, and becomes equal to the current before irradiation after 50 microstrips from the center. 

The observed deviation from the affected, due to irradiation, width of 17 microstrips, according to the beam profile, may be due to poor geometrical alignment of the detector or to other reason. This effect should be investigated. The leakage current at the central microstrip, with the exception of the three microstrips with high leakage currents, is at acceptable levels, and should have no direct effect in the pre-amplifier noise. The behaviour of the three microstrips with the high leakage current is not understood since the neighbouring strips had almost regular current. It is probable that their high current is due to combination of photoelectron induced ageing, and to mechanical stress during their mounting in the irradiation procedure.

\section*{Conclusions and prospects}

We have been able to assemble and use an experimental setup to determine the X ray  flux at the focal point of an  X-ray diffractometer . The measured fluxes were at ranges of 200 to 400~MHz, and an error in the flux of the order of $6\%$ was estimated. We have been able to use this flux of X-rays in order to study the aging of microstrip detectors for X-rays. We are investigating a method to be able to monitor even higher dosages of X-rays in facilities such as in Synchrotron radiation. 

\noindent

\section*{Acknowledgements}

We acknowledge the financial support of GSRT through the project EPET II.

\end{document}